\documentclass[conference]{IEEEtran}
\IEEEoverridecommandlockouts
\usepackage{cite}
\usepackage{amsmath,amssymb,amsfonts}
\usepackage{algorithmic}
\usepackage{graphicx}
\usepackage{textcomp}
\usepackage{xcolor}
\usepackage{float}
\usepackage[section]{placeins}
\usepackage{url}

\def\BibTeX{{\rm B\kern-.05em{\sc i\kern-.025em b}\kern-.08em
T\kern-.1667em\lower.7ex\hbox{E}\kern-.125emX}}

\begin{document}

\title{Algorithmic Design and Implementation of Unobtrusive Multistatic Serial LiDAR Image}

\author{\IEEEauthorblockN{Chi Ding, Zheng Cao, \\Matthew S. Emigh, Jos\'e C. Pr\'incipe}
\IEEEauthorblockA{\textit{Department of Electrical and Computer Engineering} \\
\textit{University of Florida}\\
Gainesville, FL\\
ding.chi@ufl.edu}
\and
\IEEEauthorblockN{Bing Ouyang, Anni Vuorenkoski\\Fraser Dalgleish, Brian Ramos, Yanjun Li}
\IEEEauthorblockA{\textit{Ocean Visibility and Optics Laboratory} \\
\textit{Harbor Branch / Florida Atlantic Univ.}\\
Fort Pierce, FL \\
BOuyang@fau.edu}
}

\maketitle

\begin{abstract}
To fully understand interactions between marine hydrokinetic (MHK) equipment and marine animals, a fast and effective monitoring system is required to capture relevant information whenever underwater animals appear. A new automated underwater imaging system composed of LiDAR (Light Detection and Ranging) imaging hardware and a scene understanding software module named Unobtrusive Multistatic Serial LiDAR Imager (UMSLI) to supervise the presence of animals near turbines. UMSLI integrates the front end LiDAR hardware and a series of software modules to achieve image preprocessing, detection, tracking, segmentation and classification in a hierarchical manner.
\end{abstract}


\section{Introduction}
We present the proposed algorithmic module of UMSLI - a Lidar-based underwater imaging system. Gaining an understanding of the underwater scene is not an easy task, especially in a visually degraded environment with low contrast, non-uniform illumination, and ubiquitous backscattering noise. It usually takes multiple modules working together to achieve the final goal. This paper discusses how these modules cooperate with each other to achieve this goal. The paper is organized as follows: the system and image data for the imaging system are discussed in Section 2. Then we dive into details of specific methods for detection, and classification in Section 3 as our main topics. Experimental results are discussed in Section 4. We will conclude the paper in Section 5.

\section{System Description}

\subsection{The Unobtrusive Multi-static Serial LiDAR Imager (UMSLI)}

The imaging system sensing front-end employs a red laser with a power density of 31.8 nJ/cm\textsuperscript{2}, which is well below 700 nJ/cm\textsuperscript{2} - the maximum permissible exposure (MPE) for human. Therefore, the system is unobtrusive and eye-safe to marine life since their eyes can focus less light than human eyes \cite{8085029}. The 638nm red laser was chosen since this wavelength is beyond their visibility range of the marine animals, and thus their behavior is not disturbed by the system. 

There are 6 fixed transmitters and receivers (cameras) deployed around the device to fully illuminate a complete 3-dimensional spherical volume so that the underwater surrounding environment can be effectively monitored. More details can be referred to from \cite{8085029}.

\subsection{LiDAR Image Data}
Compared with the conventional optical camera and imaging sonar, instead of fully explore high-resolution images or the ability to see further away, the UMSLI reaches a trade-off. There are two modes for capturing images: Sparse and Dense mode. In the sparse-mode, a pulse scan is performed by transmitters emitting red laser with lower density through a wide range, which makes the system see further and wider. The dense mode applies a high-density pulse scan for a narrower range, which gives us images with higher resolution and more detailed information within a focused area \cite{8085029}.

The data has 3 dimensions: $x$, $y$ and time $t$. $x$ and $y$ depict the pixel locations while $t$ is the time interval between the transmitter emits the laser pulse and receiver received reflected back laser signal. The range of z can be derived from t. Therefore, the signal amplitude at a spatial pixel (x,y) and at a time t represents the reflected intensity at a certain spatial point I(x,y,z). This, in turn, allows the construction of a 3-D point cloud of the scene. The advantage of this data is that it provides us with depth information and we can use this information for image preprocessing \cite{8085029}. A 2-D I(x,y) image can be formed by summing over the time axis.

Currently, our algorithms are still developed under 2D image data without using depth information. The depth information usage and automation for image restoration algorithms will be discussed in our future work. 

\section{The Algorithmic Framework}
After the image-acquisition module, a series of signal processing blocks have been developed for marine animal detection and classification. Before detection, some preprocessing techniques are applied to mitigate the typical underwater image problems such as backscattering-induced low contrast and low signal-to-noise ratio (SNR). This process enables the detection algorithm to reduce false alarms. After these steps, to achieve the final goal “classification”, tracking and segmentation methods are adopted for capturing and extracting detailed object information for our classifier. While our initial image enhancement relied on the LiDAR image enhancement technique outlined in \cite{123456}, further enhancement has been developed and implemented to achieve the desired results.

\subsection{Illumination Correction}
\begin{figure}[htbp]
\centerline{\includegraphics[scale=0.55]{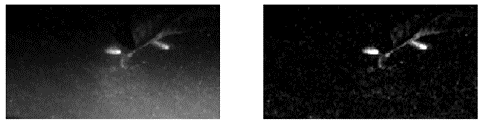}}
\caption{Result of an underwater LiDAR image. Left: original image. Right: after illumination correction.}
\label{illu_corr}
\end{figure}

When the system is operated in the sparse mode for detection, backscattering and attenuation in the turbid water are the main reasons for false alarms. As is shown in Fig.~\ref{illu_corr} left, the backscattering across the entire image can cause false detection among the area underneath the turtle with high-intensity values. It is observed that this non-uniformity is very similar to the non-uniform illumination problem. An illumination correction method was devised to correct the non-uniformity in the images \cite{illum}. Firstly, morphological open with a large structuring element $\mathbf{s}$ is applied over the image $\mathbf{I}$ to produce a background estimation $\mathbf{B}$. This step helps to remove small-scale details from the object-of-interest area and preserve background noise in $\mathbf{B}$. Then we subtract $\mathbf{B}$ from the original image to obtain an enhanced image $\mathbf{E}$.

\begin{equation}
B=I\circ s
\end{equation}

\begin{equation}
E=I-B
\end{equation}

The effectiveness of the illumination correct for the sparse-mode object detection is illustrated in Fig.~\ref{illu_corr}. However, in the dense-mode, this morphological opening method is less effective because a large structuring element has difficult to remove foreground and preserve noise when the object area is larger. A more specialized method for image restoration is required.

\subsection{Detection}
To localize salient objects, we apply a detection algorithm “gamma saliency” proposed by Burt \textit{et al.} \cite{7471948} This algorithm defines a convolutional mask “Gamma Kernel” by equation \ref{eq:gamma} and saliency map is obtained by convolving the kernel with underwater images. Gamma kernel enhances local contrast by approximating statistics of objects and the surrounding area. The kernels are displayed as “donut” shapes with a radius of approximately $\frac{k}{\mu}$ \cite{7471948}. When object size is similar to kernel size, the neighborhood is highlighted after convolution. However, if kernel radius is small, image is center-focused on each pixel. Then by subtracting neighbor-highlighted image by center-focused image, we keep high-intensity values from the object and compress areas outside of the object. To construct effective gamma kernels, we use multiple gamma kernels from equation \ref{eq:multi_gamma} with different kernel order $k$ and decay factor $\mu$ to make sure objects with different sizes can be all detected \cite{7471948}.

\begin{equation} \label{eq:gamma}
g_{k,\mu}=\frac{\mu^{k+1}}{2\pi !k} \sqrt{n_1^2 + n_2^2}^{k-1} \exp^{-\mu \sqrt{n_1^2 + n_2^2}}
\end{equation}

\begin{equation} \label{eq:multi_gamma}
g_{total}=\sum^{M-1}_{m=0} (-1)^m g_m(k_m,\mu_m)
\end{equation}

\subsection{Tracking}
After the detection, the dense-mode scanning is required to be triggered for capturing more detailed information of the detected objects. 

However, the imager performs the scene-capture via the serial scanning. The marine animal can swim further away when the imager is reading and processing data before the dense-scan is complete. Therefore, short term tracking and state prediction are necessary for guiding the system to localize position for the dense scan. Currently, the Kalman filter is applied as a preliminary step for simulation purposes. But considering the unpredictability of the real-world data and low frame rate with our system, it is highly possible a better algorithm is needed. Considering the length of this paper, the tracking algorithm will not be discussed in detail. However, we will address this issue in our future study.

\subsection{Classification}
Extensive alternatives are available from state-of-art methods such as CNN-based methods \cite{7404375}. However, due to the lack of real-world training data, traditional computer vision techniques instead of CNN-based methods are preferred at the current stage, and thus we use the information point set registration algorithm proposed in the previous work\cite{7472148}. 

This algorithm uses shape context as descriptors for the query objects. We firstly extract query shape point sets $\mathbf{X}$ and template shape point sets $\mathbf{Y}$ from threshold segmentation map. Then, shape contexts are computed from point sets $\mathbf{X}$ and $\mathbf{Y}$, they are denoted as $\mathbf{SC_x}$ and $\mathbf{SC_y}$. When classifying a query, cosine distance between shape contexts of this query and templates from each class are computed and they are noted as $\mathbf{d_{ij}}$, where $\mathbf{i}$, $\mathbf{j}$ indicates template $\mathbf{i}$ of class $\mathbf{j}$. The mean distance between this query and one specific class is calculated by averaging the sum of $\mathbf{d_{ij}}$ with respect to $\mathbf{i}$ for class $\mathbf{j}$. Then the query is assigned to the class $\mathbf{j}$ which has the least average distance.

\begin{equation}
d_{ij}= \frac{1-SC_{X}\cdot SC_{Y_{i,j}}}{\lVert X\lVert \cdot \lVert Y_{i,j}\lVert}
\end{equation}

\begin{equation}
\bar{d_{j}} = \frac{\sum_{i=1}^{n_{j}}d_{ij}}{n_{j}}
\end{equation}

Usually using the descriptor alone is unable to achieve high classification accuracy because query shapes can be very noisy or distorted in real-world conditions. The noise and distortion can be caused by backscattering, partial bodies or variations due to the different poses of the object. To make classifier more robust, we introduce a similarity measure between two aligned shapes. A projection matrix $\mathbf{A}$ is learned within a certain number of iterations by maximizing the similarity of $\mathbf{XA}$ and template point set $\mathbf{Y}$ under maximum correntropy criterion (MCC)\cite{JMLR:v16:feng15a}. In each iteration, $\mathbf{X}$ is updated by $\mathbf{XA}$. By doing this, the query point set $\mathbf{X}$ performs an affine projection by multiplying matrix $\mathbf{A}$ in each iteration that enables the query shape to align with the template. Because each time $\mathbf{X}$ performs an affine transformation, the correntropy (similarity) between itself and the template $\mathbf{Y}$ are maximized. Therefore, we calculate the correntropy similarity measure $\mathbf{c_j}$ between the aligned query and the template. The final dissimilarity score $\mathbf{D_j}$ is calculated by dividing $\mathbf{d_j}$ by $\mathbf{c_j}$. Then we assign query $\mathbf{X}$ to the class $\mathbf{j}$ with minimum average dissimilarity measure $\mathbf{\bar{d_j}}$ across all other classes. 

\begin{equation}
c_j=Corr(XA,Y_j)
\end{equation}

\begin{equation}
\bar{d_j} = \frac{\sum_{i=1}^{n_j}\frac{d_{ij}}{c_{ij}}}{n_j}
\end{equation}

The advantage of this method is that correntropy captures the higher-order statistic information and it is robust against noise \cite{7472148, JMLR:v16:feng15a, 4355325}. These properties guarantee efficient alignment when two objects are similar or dissimilar.

Another advantage of this method is that there are many other good shape descriptors to choose from, they can be mostly classified into two categories: area-based and boundary-based (Shape context is one of them). Boundary based methods are usually rich in representing details of shapes while area-based methods are more robust to noise. Therefore complementary shape descriptors can be chosen and integrated into our classifier according to the varying real-world conditions. According to our previous work, the shape can also be defined as a single ``view'' and integrated with other views (features) \cite{8444645}, such as texture, to improve classification accuracy.

Furthermore, compared with the most popular state-of-art CNN-based methods, this algorithm does not require large amount of training data or a very complicated model, which is hard to update if new classes/animals appear. This method gives us the flexibility of building a new model or adapting the existing model very quickly with smaller efforts.

\subsection{Reinforcement Learning Template Selection: Divergence to Go}
However, it can be troublesome when we have too many instances/templates, especially many of them are highly similar. A correct classification result can always be attained by comparing with as many shape templates as possible if they follow the correct distributions within their own class, but computation cost can be very high. Therefore, we apply the “divergence-to-go” (DTG) reinforcement learning framework for selecting shape templates to reduce the amount of computation load for our classifier. 

Since this is a classification task, the templates should be as discriminative as possible over other classes. For example, a template for amberjack should look very different from a template for barracuda. To achieve this goal, we set up a reinforcement learning experiment applying the DTG policy, which maximizes the uncertainty over the next step, to transit from one state (Hu moments \cite{1057692} of a template within one species) to another. The training process takes certain iterations and we keep track of the number of visits for each state. Then we select the n most visited shapes as they preserve the most uncertainty with respect to other classes.

DTG is a quantity that measures the uncertainty given a state-action pair and defined as the expected discounted sum of divergence over time \cite{7324371}. This framework is similar to a Markov Decision Process (MDP) with the 5-tuple ($\chi, A, P, R, \gamma$) substituting reward $R$ as uncertainty, which is measured by divergence that calculates the distance between two transition density distributions. Transition distribution and divergence are estimated using kernel density estimators under the kernel temporal difference (KTD) model \cite{6064634}.

\begin{equation}
dtg(x,a)=E\left[ \sum^{\infty}_{t=0} \gamma^{t} D(x_t) \right]
\end{equation}

Applying the dynamic programming framework for temporal credit assignment and we have the DTG update equation: 
\begin{equation}
\delta_t=D+\gamma \max_a \{dtg_{t+1}(\bar{x^{'}})\}-dtg_t(\bar{x})
\end{equation}

\begin{equation}
dtg_t(\bar{x})=D_0+\alpha \sum^{t}_{j=1}\delta_{j} k(\bar{x},\bar{x_j}),
\end{equation}where $\mathbf{D}$ is divergence measure, $\mathbf{\delta_t}$ is DTG error and $k(.)$ is a similarity function of two states, here correntropy \cite{4355325} is adopted. 

\section{Experiments and Results}
\subsection{Description of the Test Dataset}
Two different datasets were used in this study. In November 2017, a Gen-I UMSLI prototype was deployed in the DOE test facility operated by the Marine Sciences Laboratory at Pacific Northwest National Laboratory (PNNL). During the two-week exercise, a substantial amount of images were acquired using the UMSLI prototype. These include images of the two artificial targets (turtle and barracuda) and images of harbor seals and other natural fishes.

\begin{figure}[htbp]
\centerline{\includegraphics[scale=1]{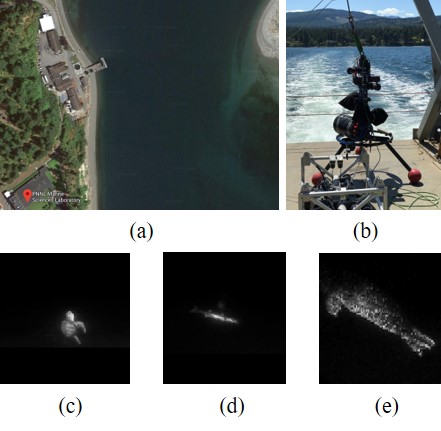}}
\caption{Illustration of the PNNL/MSL Test site, instrument and sample test images acquired during the experiments. (a) PNNL/MSL test site (b) GEN-I UMSLI prototype (c) Artificial turtle (d) Artificial baracudda (e) Natual harbor seal}
\label{dataset}
\end{figure}

In addition, a bench-top system of Gen-II UMSLI with the improved optical and electronic system was used to validate the automated switching from sparse-mode scanning and dense-mode scanning after detection.

\subsection{Detection}

We compare the performance of gamma saliency and 6 other most highly cited state-of-art methods based on detection datasets acquired at DOE PNNL/MSL test site in this study. These methods are Hou and Zhang\cite{4270292}, B. Schauerte \textit{et al.}\cite{Schauerte:2012:QSS:2403006.2403016}, Achanta \textit{et al.}\cite{5206596}, Margolin \textit{et al.} \cite{6618995}, Goferman \textit{et al.} \cite{6112774}, and Jiang \textit{et al.}\cite{6751317}, denoted as SR, QT, IG, PCA, CA, and MC respectively. Reasons for choosing these six methods are twofold: 

\textbf{\emph{Diversity:}} PCA, CA are complicated region based methods\cite{Borji2014SalientOD}. 
PCA computes the distance between every patch and the average patch in PCA coordinates to maximize inter-class variations. CA incorporates both single-scale local saliency 
measured by surrounding patches and multi-scale global contrast information.
MC is a segmentation method based on superpixels and absorbing Markov chain. Saliency is computed by weighted sum of absorbed time from transient state to the absorbed state.
SR and QT are spectral based fast models by looking at salient regions as residual information in image spectral space by using Fourier Transform and Quaternion Fourier Transform respectively. IG uses band-pass filter to contain a wide range of frequencies information and then finds the saliency 
map by subtracting filtered image by the arithmetic mean pixel value to get rid of texture details.

\textbf{\emph{Speed Limit:}} PCA, CA are computationally extensive while SR, QT, IG, and MC are fast.

To measure performance of all methods, four widely-used and universally agreed metrics are 
applied for comparison purpose: 
\begin{enumerate}
\item \textbf{\emph{Precision and Recall (PR) Curve}}
\item \textbf{\emph{Receiver Operating Characteristic (ROC) Curve}}
\item \textbf{\emph{F-measure}}
\item \textbf{\emph{Area Under Curve (AUC)}}
\end{enumerate}

Before calculating the PR curve, saliency 
maps are binarized with threshold values from 0 to 255 to get the binary masks\textbf{\emph{M}}. 
Then precision and recall are calculated by comparing the binary mask \textbf{\emph{M}}
with the ground truth \textbf{\emph{G}}. PR curve evaluates overall performance in terms of positive classes while ROC curve evaluates both positive and negative classes. For calculating F-measure, according to the adaptive binarization method 
Achanta \textit{et al.} proposed \cite{5206596}, adaptive threshold value is chosen as parameter $\alpha$ times the mean value of saliency map. Instead of using fixed $\alpha$, we evaluate the overall F-measure by using values from 2 to 11 to realize fair comparison, because saliency maps based on different algorithms are sensitive to different adaptive threshold values. Then all F-measure values given different $\alpha$ are averaged as our final result as is shown in table 1.
\begin{equation}
F_{\beta} = \frac{(1+\beta^{2})Precision\times Recall}{\beta^{2}Precision + Recall}
\end{equation}
According to \cite{5206596}, $\beta^{2} = 0.3$ is selected since precision is more important than recall as high recall can be easily achieved by lowering threshold value so that binary mask mostly covers ground truth. 

\begin{figure*}[htb]
\includegraphics[width=\textwidth,height=5cm]{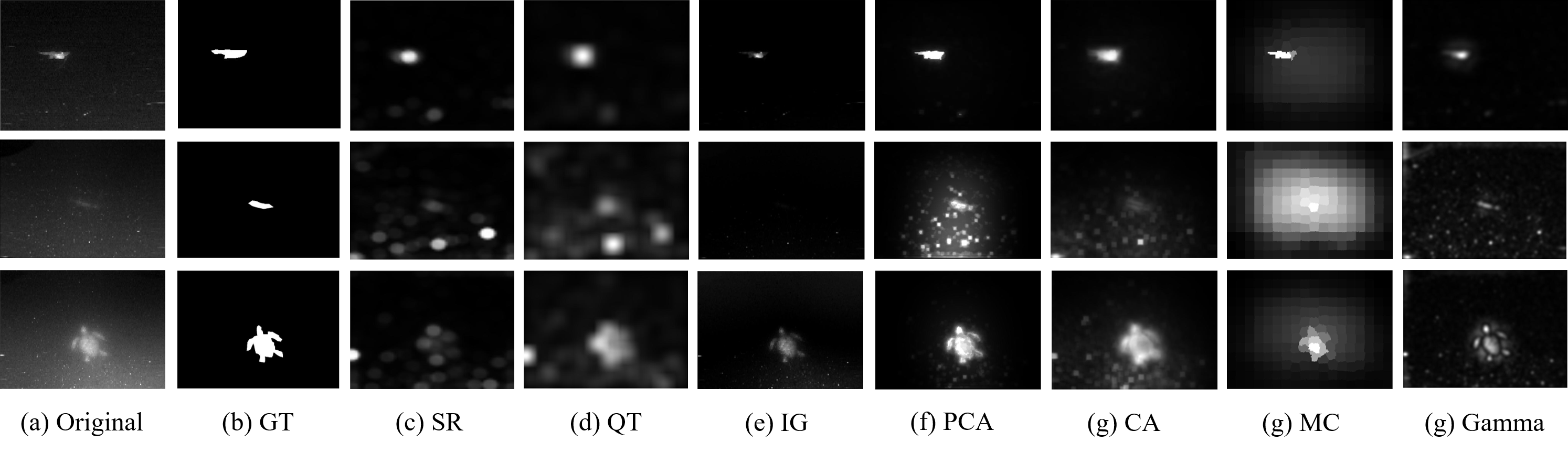}
\caption{\textit{Comparison of saliency map results. From left to right: 
Original image, ground truth, SR, QT, IG, PCA, CA, MC, gamma.}}
\label{fig_result}
\end{figure*}

Qualitative results of saliency detection on a natural small fish, artificial barracuda and turtle models are shown in Fig.~\ref{fig_result}.

\begin{figure}[htbp]
\centerline{\includegraphics[scale=0.18]{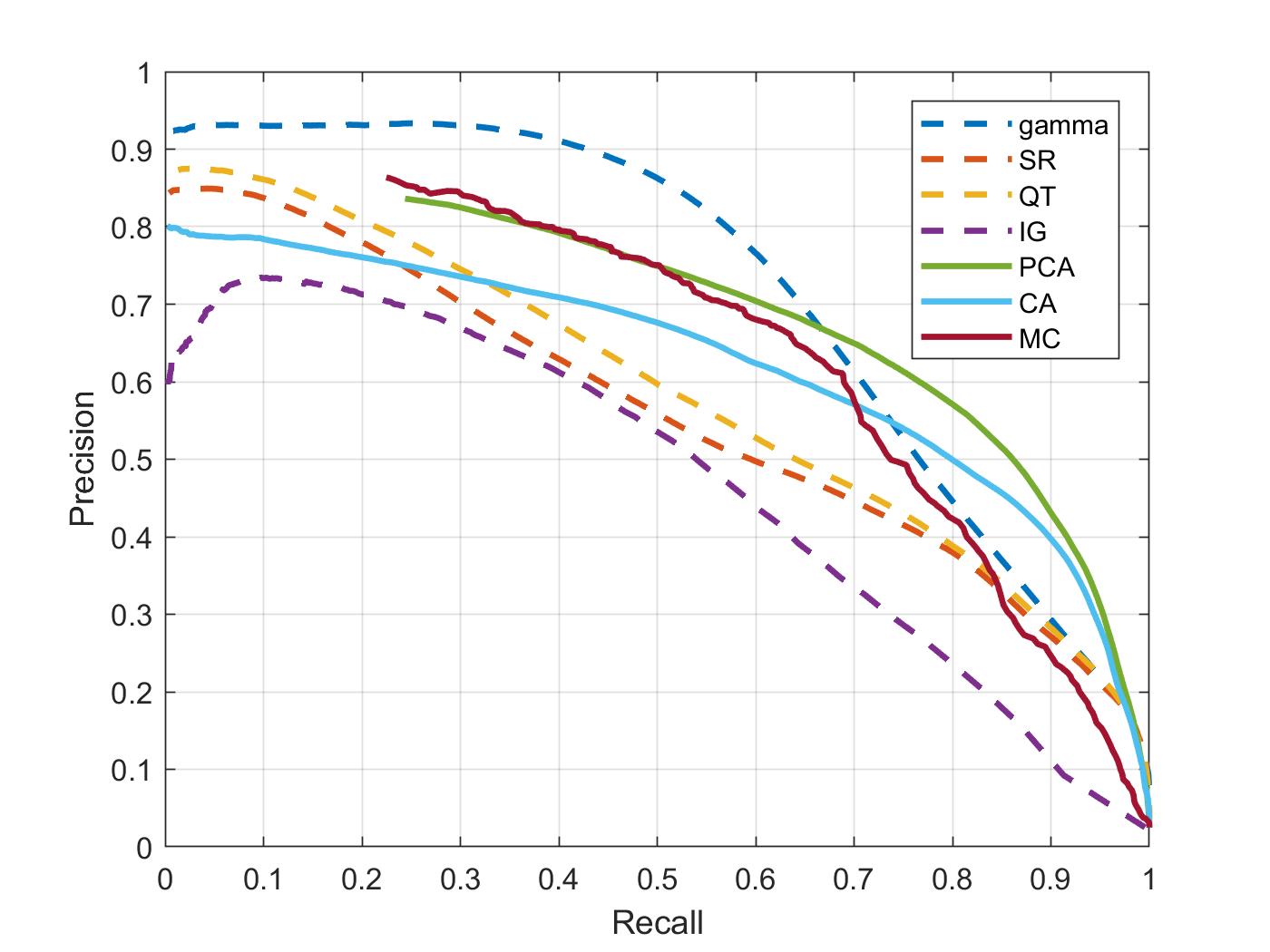}}
\caption{PR curve. Comparisons for performances of gamma saliency 
against six methods: SR, QT, IG, PCA, CA, MC. Fast algorithms are plotted 
as dashed lines to seperate from comutationally extensive algorithms.}
\label{PR}
\end{figure}

In Fig.~\ref{PR}, gamma saliency outperforms all the fast algorithms but not PCA or CA, which means gamma saliency demonstrates better accuracy than fast algorithms but produces more false positives than computational extensive region-based models. However, when the recall value is less than 0.65, gamma dominates all other methods. This result implies that gamma is useful for object localization without considering the fine details. This is demonstrated in Fig~\ref{fig_result} that gamma only preserves saliency blob. According to the nature of our application, localization is the most important task before segmentation and classification, which can be achieved by gamma saliency efficiently.

\begin{figure}[htbp]
\centerline{\includegraphics[scale=0.18]{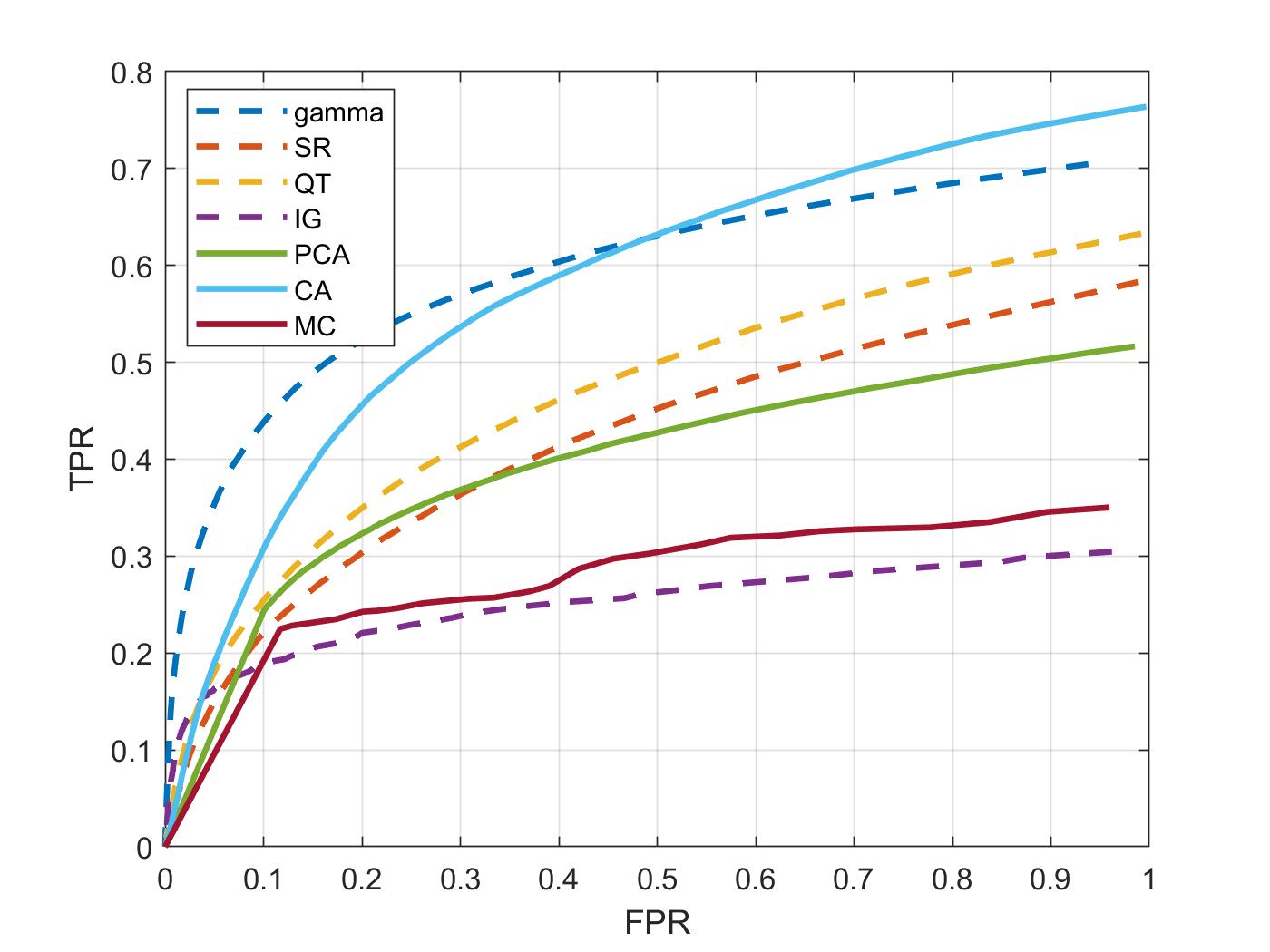}}
\caption{ROC curve. Comparisons for performances of gamma saliency 
against six methods: SR, QT, IG, PCA, CA, MC. Fast algorithms are plotted 
as dashed lines to seperate from comutationally extensive algorithms.}
\label{ROC}
\end{figure}

According to the ROC curve in Fig.~\ref{ROC}, the gamma saliency outperforms other methods. The intersection of gamma saliency and CA around 0.5 FPR indicates that gamma captures more saliency than CA in terms of regions other than the object of interest because gamma kernel picks up wherever saliency appeared while CA integrates both local and global contrast information\cite{6112774}. Therefore, a higher threshold value is needed for gamma to approach better quality saliency map. Even though CA performs better than gamma saliency, CA cannot meet the requirement for real-time saliency detection. More details are shown in table~\ref{gamma_table}.

The area-under-curve (AUC) is also provided in Table 1 to validate and support our conclusion that gamma saliency maintains an overall good performance with low computation time since the AUC value of gamma is higher than all fast methods and slightly lower than computational extensive methods.

Saliency maps for three representative images of different methods are provided in Fig.~\ref{fig_result} The first row is a natural fish captured by our underwater imager, the second row is the artificial barracuda model appears in the center of the image with extremely low contrast (in the far-field), and the third row is the artificial turtle model in turbid water. 

As shown in the first row of Fig.~\ref{fig_result}, all methods capture the fish accurately under clear water background even though the object is small. However, when contrast between foreground and background is low, other methods either recognize noise as salient region (SR, QT) or unable to distinctively separate noise from object (PCA, MC). In the worst-case, IG cannot distinguish between the salient region and background at all. CA and gamma detect the object but only gamma successfully separate object from a noisy background.

In conclusion, gamma satisfies the following properties for detection in a real-time manner:
\begin{itemize}
\item Localize the whole salient object with high accuracy.
\item High robustness against noise.
\item Computationally efficient.
\end{itemize}

\begin{table*}[ht]
\caption{Comparison}
\centering
\begin{tabular}{c c c c c c c c c c}
\hline\hline
Measure & SR & QT & IG & PCA & CA & MC & Gamma & Avg\\ [0.5ex]
\hline
F-measure & 0.4890 & 0.4978 & 0.5393 & 0.6357 & 0.5559 & \textbf{0.6463} & 0.5932 & 0.5653\\
AUC & 0.3486 & 0.4594 & 0.2400 & 0.3868 & \textbf{0.5755} & 0.2640 & 0.5505 & 0.4035\\
Time (s) & \textbf{0.0464} & 0.1481 & 0.2229 & 11.5690 & 19.6659 & 0.5348 & 0.4012 & 4.6554\\ [1ex]
% content
\hline
\end{tabular}
\label{gamma_table}
\end{table*}

\subsection{Classification}
\begin{figure}[H]
\centerline{\includegraphics[scale=0.56]{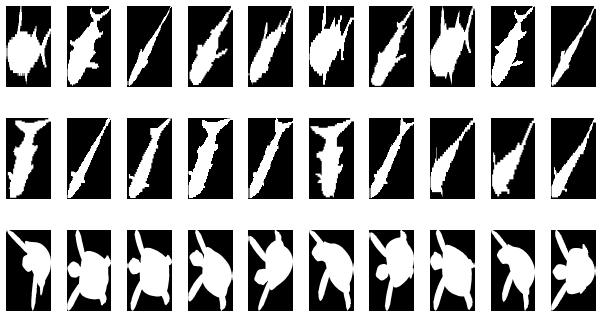}}
\caption{10 templates chosen by k-means clustering for each classes. First row: amberjack, second row: barracuda, third row: turtle \cite{Cao2017}.}
\label{kmeans}
\end{figure}

There are 256 instances for each class generated by projecting a 3D shape model into 2D from different angles, and they are used as shape template sets $\mathbf{Y}$ as mentioned before. The experiment for classification is related to $DTG$ policy selecting templates. We select 10 templates by DTG policy and compare its performance with k-means clustering from previous work \cite{Cao2017}. Because each class has 256 templates, the previous method calculates $256\times256$ similarity matrix and each row represents one shape template. Then k-means clustering is applied to cluster all templates into 10 categories, the representative of each category is simply selected by the one shape vector that has the least L1-norm \cite{Cao2017}. The selection is based on the idea of fully representing the class with fewer instances, which gives reasonably good results for a small data set. However, discrimination between classes is not included under such scenario. In Fig.~\ref{kmeans}, it is obvious that templates for turtle are very different from both barracuda and amberjack. However, there are clear similarities between templates of barracuda and amberjack, such as the first-row third-column and second-row second-column in Fig.~\ref{kmeans}. Therefore, it is necessary that discrimination among different classes are introduced into the proposed method for selecting templates.

DTG policy is a model-based reinforcement learning method, we can achieve the task of introducing discrimination by manipulating the transition model. Simply saying, if we want to train a policy that maximizes the uncertainty (divergence) of amberjack with respect to barracuda, then we train the DTG policy of amberjack based on the transition model of barracudas. We calculate Hu-moments for all shape templates as their corresponding states. Hu-moments can effectively discourage dissimilarity caused by translation, scale, rotation, and reflection \cite{1057692}. These properties are important because many templates are highly similar to each other even though they are shifted, rotated or translated and applying Hu-moments can discard these variations and focus more on inherent information of shape itself. 

The experiment set-up is given as follows: Firstly calculating similarity matrix for all three classes that we have: amberjack, baracuda, and turtle. Then we apply k-means to cluster each class into 10 clusterings respectively and select 20 representatives that have the least L1-norm from the 10 clusterings. Then we define action list [-10, ..., -1, 1, 2, … , 10] ($0$ is eliminated because staying at the same state should be avoided for DTG policy). Each action $\mathbf{a}$ transits state $\mathbf{i}$ into state $(a + i)$ $mod$ $20$. The transition model is built by running this environment by random policy for 5000 steps and then storing all transitions: [$x_i$, $x_{i+1}$, $a$, $r$]. After building transition models for all classes, we start training DTG policies within certain steps. Then we narrow down the 20 templates into 10 by selecting the templates that correspond to the 10 most highly visited states within the 2,000 steps with respect to other classes. We compare the performance of templates selected by DTG and the original k-means method as well as random policy. The original k-means method calculates confusion matrix based on previous dataset composed of 8 amberjacks, 6 barracuda and 8 turtles \cite{Cao2017}. Because this dataset is too small, we also compare the classification accuracy based on the same dataset for detection task, which only has 2 classes: barracuda and turtle (because templates for other natural fishes we encountered are not available). For the comparing experiment based on 3 classes, DTG policy of one class is trained under the concatenation of all transition models of other classes. For the experiment involving the only 2 classes, the DTG policy is only trained under the transition model of the other class.

\begin{figure}[htbp]
\centerline{\includegraphics[scale=.45]{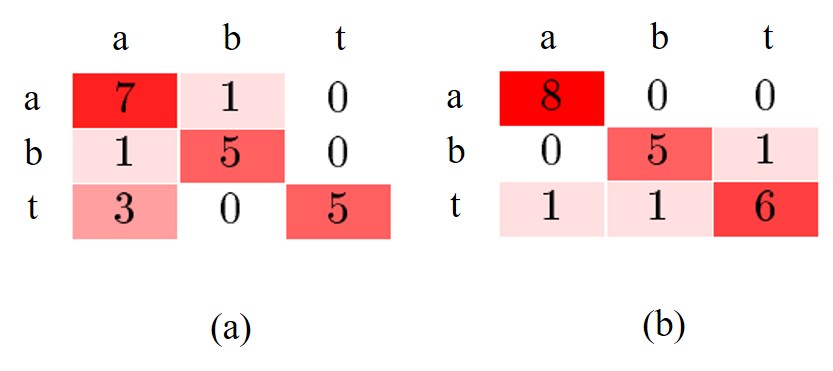}}
\caption{Confusion matrix of classification result given by temlates selected by different methods where a, b and t stands for amberjack, barracuda and turtle respectively. (a) k-means (b) DTG}
\label{confusion_matrix}
\end{figure}

Classification results for two different dataset are provided in Fig.~\ref{confusion_matrix} and Table~\ref{rate} respectively. In the confusion matrix, it is clearly demonstrated that discrimination is introduced into template selection since there are fewer false positives and more true positives. Classification accuracy is the highest among all methods that we apply for templates selection.

\begin{table}[ht]
\caption{Classification Rate} % title of Table
\centering % used for centering table
\begin{tabular}{c c c c} % centered columns (4 columns)
\hline\hline %inserts double horizontal lines
& DTG & Random & K-means \\ [0.5ex] % inserts table
%heading
\hline % inserts single horizontal line
Barracuda & \textbf{93.99}\% & 87.24\% & 90\% \\ % inserting body of the table
Turtle & \textbf{97.5\%} & 93.00\% & 92.5\% \\ [1ex] % [1ex] adds vertical space
\hline %inserts single line
\end{tabular}
\label{rate} % is used to refer this table in the text
\end{table}

\subsection{Camera System Implementation Results}
\begin{figure}[htbp]
\centerline{\includegraphics[scale=0.48]{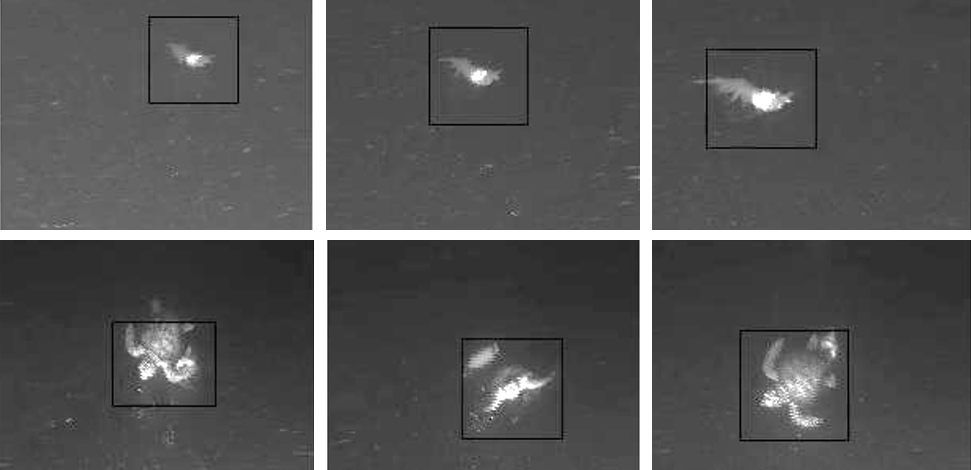}}
\caption{Detection results on real-world data.}
\label{real_world}
\end{figure}

\begin{figure}[htbp]
\centerline{\includegraphics[scale=0.9]{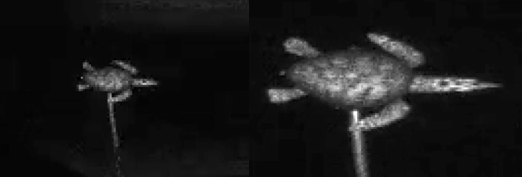}}
\caption{Real-time automatic switching from sparse-mode to dense mode after the object is detected: more details (texture, object shape) are available.}
\label{switch}
\end{figure}

Currently, the detection algorithm has been implemented in the embedded system to realize real-time automatic switching from the sparse-mode to the dense-mode when an object is detected. A series of experiments were conducted at the optical test tank at HBOI to validate this implementation. Offline detection results are shown in Fig.~\ref{real_world} and real-time mode switching results are shown in Fig.~\ref{switch} 

For real time object detection and mode switching, we hide bounding box for acquiring clear data.

\section*{Conclusion}
First of all, using the field data acquired at PNNL/MSL test site, we demonstrated both quantitatively and qualitatively that the optimality of the gamma saliency algorithm in real-time detection of undersea animals. One critical step in achieving this success is that the illumination correction based image processing was able to mitigate backscattering and attention induced image distortion that is typical in the turbid water.

Furthermore, the gamma saliency algorithm has been integrated into the GEN-II UMSLI benchtop prototype to demonstrate the ability of automatic switching from the sparse-scan mode to the dense-scan mode – critical for the comparatively low-frame-rate UMSLI imager to successfully acquire the high-resolution images needed for the subsequent classification task.

The DTG reinforcement learning-based template selection method introduced in this paper was based on the belief that templates representing a class should be both representative and discriminative. The simulation results demonstrate the effectiveness of this technique. The experiments for classification demonstrated initial success in developing a highly flexible shape-matching framework that can evolve to incorporate more features \cite{8444645} or additional templates to when additional data become available. This ability is critical to the success of the UMSLI deployment at any MHK site where the encounters with underwater animals are in general scarce. 

\section*{Acknowledgement}
This work was partially supported by USDOE contracts DE-EE0006787 and DE-EE0007828. The authors will also want to thank Mr. Michael Young for fabricating the system components and engineers at PNNL/MSL: Dr. Genevra Harker-Klimes, Mr. Garrett Staines, Mr. Stanley Tomich, Mr. John Vavrinec and Ms. Shon Zimmerman for their support. 

%\bibliographystyle{IEEEbib}

%\bibliography{Itti,KTD,illum,MCC,enhance,Hu,dtg,multiview,zheng_cnn,zheng_shape,corr,SR,QT,QT1,PCA,CA,MC,distinct,unique,object,LiDAR,gamma_saliency,NN_gamma,gamma,zheng,IG,survey}

\end{document}